\documentclass[prd,aps,12pt,floatfix]{revtex4}
\usepackage{epsfig}
\begin{document}

\title{Twisted mass QCD for the pion electromagnetic form factor}

\author{Abdou M. Abdel-Rehim}

\author{Randy Lewis}

\affiliation{Department of Physics, University of Regina, Regina, SK,
S4S 0A2, Canada}

\begin{abstract}
The pion form factor is computed using quenched twisted mass QCD and the
GMRES-DR matrix inverter.  The momentum averaging procedure of Frezzotti and
Rossi is used to remove leading lattice spacing artifacts, and numerical
results for the form factor
show the expected improvement with respect to the standard Wilson action.
Although some matrix inverters are known to fail when applied to twisted
mass QCD, GMRES-DR is found to be a viable and powerful option.
Results obtained for the pion form factor are consistent
with the published results from other $O(a)$ improved actions and are also
consistent with the available experimental data.
\end{abstract}

\maketitle

\section{Overview and discussion}

The Wilson lattice action for fermions\cite{Wilson}
does not have exact chiral symmetry,
and thus it provides no lower bound for the norm of Dirac matrix eigenvalues.
As a practical consequence, numerical simulations encounter
``exceptional configurations'' (for which the Dirac matrix cannot be inverted)
with increasing regularity as the quark mass is reduced.  Symanzik improvement
of the Wilson action by adding the Sheikholeslami-Wohlert term with a properly
tuned coefficient\cite{clover}, exacerbates the problem.

Fortunately,
the addition of a twisted mass term provides a lower bound for the eigenvalues
thereby eliminating exceptional configurations.\cite{tmQCD,FreLat04}
Furthermore, tuning of the Sheikholeslami-Wohlert coefficient is no longer
needed for the removal of linear lattice spacing errors.  Instead, many
quantities are automatically improved simply by adding the twisted
mass term to the basic Wilson action and setting the hopping parameter to its
critical value (i.e. $\kappa=\kappa_c$ in Eq.~(\ref{tmQCD-lattice-action})
below), while other quantities become improved by
averaging over equal and opposite momenta.\cite{FreRos,FreLat04}
The present work provides an explicit
numerical verification of the averaging technique for improvement in
twisted mass QCD (tmQCD).

Our discussion will focus on the pion electromagnetic form factor.
Due to the simple valence quark structure of the pion and a firm theoretical
knowledge of its form factor in both the $Q^2\to0$ and $Q^2\to\infty$ limits,
the pion form factor is a preferred place to study the transition between
perturbative and nonperturbative QCD.  Experimental studies are ongoing
at Jefferson Lab, and theoretical modelling is also continuing.\cite{experiment}

Initial studies of the pion form factor using lattice QCD occurred some time
ago,\cite{firstlat,Draper} and new lattice initiatives have arisen recently
for Wilson, Sheikoleslami-Wohlert, and domain wall
actions.\cite{LHPC,vanderH,Nemoto}
Preliminary results from our tmQCD study were presented in
Ref.~\cite{ourprelim}.  In contrast to other
form factors, the pion form factor receives no contributions from the vector
current attaching to a nonvalence quark (so-called ``disconnected diagrams''),
and this feature reduces the lattice QCD cost considerably.\cite{Draper}
There could still be contributions from sea quarks that do not interact
directly with
the external vector current, and these have been considered in
Ref.~\cite{dynapion} where dynamical configurations were to used obtain the
pion form factor.  All other studies to date, including the present one,
have used the quenched approximation and thereby omitted all nonvalence quarks.

In the remainder of this article, we report on our use of quenched tmQCD to
compute the pion form factor.  Two quark masses corresponding to pion masses
near 470 MeV and 660 MeV, as well as a variety of momentum
transfer values satisfying 0 GeV$^2<Q^2\lesssim5$ GeV$^2$ have been considered,
all at a lattice spacing of 0.10 fm.
A comparison to existing lattice results clearly shows the improvement
expected for tmQCD, since the momentum-averaged tmQCD data agree with
results from other improved actions and differ from unimproved Wilson results
at this same lattice spacing.
Interestingly, even before momentum averaging the tmQCD data are closer to
improved action results than to unimproved Wilson results for this particular
observable, despite the fact that the pion form factor technically requires
momentum averaging to exactly remove the linear lattice spacing errors.

To determine the renormalization factors that appear in the pion form factor
correlation function, and to compare with the predictions of vector meson
dominance, we also study two-point pseudoscalar and vector
correlators with nonzero momenta.  The associated dispersion relations
are compared to continuum expectations as another means of exploring lattice
spacing artifacts.

One of the technical issues that arises in tmQCD simulations is the failure
of some standard matrix inversion algorithms.  Alternative algorithms are
being used and evaluated by various authors.\cite{inverters}
The present work makes use of the GMRES-DR algorithm\cite{GMRESDR}
and concludes that it performs well for tmQCD.
Some details are presented in Section~\ref{sec:gmresdr}.

This initial exploration of the pion form factor with tmQCD leads to optimism
that future lattice tmQCD studies, perhaps with smeared operators and increased
statistics on larger lattices, can reach smaller quark masses
with greater precision.  Our present results are consistent with
vector meson dominance and with experiment.
More generally, the present work underscores the value of lattice tmQCD itself
as a practical tool for hadron phenomenology.

\section{Correlation functions}

The electromagnetic form factor of a charged pion is defined by
\begin{equation}
\label{piff-matrix-element}
\left<\pi^+(\vec{p}_f)|j_\mu(0)|\pi^+(\vec{p}_i)\right>=F(Q^2)(p_i+p_f)_\mu
\end{equation}
where $j_\mu(0)$ is a conserved vector current evaluated at the spacetime
origin, $p_i$ and $p_f$ are the initial and final pion (Euclidean) 4-momenta 
respectively, $\vec p_i$ and $\vec p_f$ are the corresponding 3-momenta,
and $Q^2=(p_f-p_i)^2$ is the 4-momentum transfer.  To compute
this matrix element on a spacetime lattice, one can use the three-point
correlator displayed in Fig.~\ref{3point}.
A source with pion quantum numbers is placed at $x_i$, a sink at $x_f$,
and a vector current is inserted at $x$.
Given an interpolating field operator $\phi(x)$ with the quantum numbers of
the charged pion, one can extract the form factor from the following
three-point correlator:
\begin{equation}
\label{3pt-correlator-expression}
\Gamma_{\pi\mu\pi}(t_i,t,t_f,\vec p_i,\vec p_f)=
\sum_{\vec x_i,\vec x_f}
e^{-i(\vec x_f-\vec x)\cdot\vec p_f}
e^{-i(\vec x-\vec x_i)\cdot\vec p_i}
\left<0|\phi(x_f)j_\mu(x)\phi^\dagger(x_i)|0\right>.
\end{equation}
Note that we are working in units of lattice spacing throughout this
discussion.  In this work we choose $\phi(x)$ to be the local operator
\begin{equation}
\phi(x)=\bar{d}(x)\gamma_5u(x)
\end{equation}
where $u(x)$ and $d(x)$ are the up and down quark fields respectively.
Smeared operators could be of value in subsequent studies, particularly
for the exploration of the high $Q^2$ range.
For the vector current, $j_\mu(x)$, we use the conserved current,
\begin{equation}
j_\mu(x) = \frac{1}{2}\bar{u}(x)(1-\gamma_\mu)U_\mu(x)u(x+\hat{\mu})
         - \frac{1}{2}\bar{u}(x+\hat\mu)(1+\gamma_\mu)U_\mu^\dagger(x)u(x)
\end{equation}
with $\mu=4$.
In order to extract the matrix element
in Eq.~(\ref{piff-matrix-element}) from the three-point correlator of
Eq.~(\ref{3pt-correlator-expression}),
one introduces two complete sets of states $|n(\vec{k})\rangle$ with the 
same quantum numbers as $\phi(x)$ in the three-point correlator and gets
\begin{eqnarray}
\Gamma_{\pi\mu\pi}(t_i,t,t_f,\vec p_i,\vec p_f)
&=&\sum_n \sum_m \left<0|\phi(x)|m(\vec{p}_f)\right>
   {{e^{-(t_f-t)E_m(\vec{p}_f)}}\over{2E_m(\vec{p}_f)}}\left<m(\vec{p}_f)
   |j_\mu(x)|n(\vec{p}_i)\right> \nonumber \\
&& {{e^{-(t-t_i)E_n(\vec{p}_i)}}\over{2E_n(\vec{p}_i)}}\left<n(\vec{p}_i)
   |\phi^\dagger(x)|0\right>.
\end{eqnarray}
This can be simplified further using 
\begin{equation}
\left<0|\phi(x)|m(\vec{p})\right>=Z_m(\vec{p})e^{ix\cdot p}
\end{equation}
and for a local interpolating field operator, $Z_m(\vec{p})$ is independent
of $\vec{p}$. The three-point correlator simplifies to
\begin{equation}
\Gamma_{\pi\mu\pi}(t_i,t,t_f,\vec p_i,\vec p_f)=\sum_n \sum_m Z_m
{{e^{-(t_f-t)E_m(\vec{p}_f)}}\over{2E_m(\vec{p}_f)}}
\left<m(\vec{p}_f)|j_\mu(0)|n(\vec{p}_i)\right> 
{{e^{-(t-t_i)E_n(\vec{p}_i)}}\over{2E_n(\vec{p}_i)}}Z_n^*.
\label{3pt-decomposition}
\end{equation}
Similarly the two-point correlator, which will be needed to get
the energies, is given by
\begin{equation}
G_{\pi\pi}(t_i,t,\vec{p}) = \sum_{\vec x}e^{-i(\vec x-\vec x_i)\cdot\vec{p}} 
\left<0|\phi(x)\phi^\dagger(x_i)|0\right>
=\sum_n Z_nZ_n^*{{e^{-(t-t_i)E_n(\vec{p})}}\over{2E_n(\vec{p})}}.
\label{2pt-decomposition}
\end{equation}
For periodic boundary conditions on a lattice of $N_t$ time slices,
Eq.~(\ref{2pt-decomposition}) will be modified to
\begin{equation}
G_{\pi\pi}(t_i,t,\vec{p}) = 
\sum_n \frac{Z_nZ_n^*}{E_n(\vec{p})}e^{-\left(\frac{N_t}{2}
\right)E_n(\vec{p})}
\cosh\left[\left(t-t_i-\frac{N_t}{2}\right)E_n(\vec{p})\right].
\label{2pt-decomposition-periodic}
\end{equation}
The long time behaviour of the two and three-point correlators will be
dominated by contributions from the lightest pseudoscalar state, i.e. the pion.
This asymptotic behaviour is given by
\begin{eqnarray}
\label{asymptotic-correlators1}
\Gamma_{\pi\mu\pi}(t_i,t,t_f,\vec p_i,\vec p_f)
& \stackrel{t_f \gg t \gg t_i}{\longrightarrow} &
|Z_\pi|^2
\frac{e^{-(t-t_i)E_\pi(\vec{p}_i)-(t_f-t)E_\pi(\vec{p}_f)}}
{4E_\pi(\vec{p}_i)E_\pi(\vec{p}_f)}F(Q^2)(p_i+p_f)_\mu, \\
G_{\pi\pi}(t_i,t,\vec p) & \stackrel{|t-t_i|\gg0}{\longrightarrow} &
\frac{|Z_\pi|^2}{E_\pi(\vec{p})}
e^{-\left(\frac{N_t}{2}\right)E_\pi(\vec{p})}
\cosh\left[\left(t-t_i-\frac{N_t}{2}\right)E_\pi(\vec{p})\right].
\label{asymptotic-correlators2}
\end{eqnarray}
To obtain a reliable result for the pion form factor, we will
allow for contributions from excited states.
For the conserved current, the corresponding transition matrix elements are
included as follows,
\begin{equation}\label{transition}
\left<\pi_\beta(\vec{p}_f)|j_\mu(0)|\pi_\alpha(\vec{p}_i)\right>
= F_{\alpha\beta}(Q^2)\left[(p_f+p_i)_\mu-\frac{(p_f^2-p_i^2)}{(p_f-p_i)^2}
(p_f-p_i)_\mu\right].
\end{equation}

\section{The action and its parameters}

Our simulations use the standard Wilson gauge action with $\beta=6.0$.
An ensemble of 100 quenched configurations of size $16^3\times48$ was
created using a pseudo-heatbath algorithm, with 5000 sweeps omitted between
saved configurations.
The lattice tmQCD fermion action is
\begin{equation}
S_F[\psi,\bar\psi,U] = \sum_x\bar\psi(x)\left[1+2\kappa\mu i\gamma_5\tau_3
+\kappa\sum_\nu\gamma_\nu\left(\nabla_\nu+\nabla^*_\nu-\nabla^*_\nu\nabla_\nu
\right)\right]\psi(x)
\label{tmQCD-lattice-action}
\end{equation}
where the forward and backward lattice derivatives are defined as usual,
\begin{eqnarray}
\nabla_\nu \psi(x) &\equiv& U_\nu(x)\psi(x+\hat{\nu})-\psi(x), \\
\nabla_\nu^*\psi(x) &\equiv&
\psi(x)-U_\nu^\dagger(x-\hat{\nu})\psi(x-\hat{\nu}),
\end{eqnarray} 
and $\psi(x)$ denotes the doublet of up and down quarks.
When $\mu=0$, $S_F$ becomes the standard Wilson action.
Throughout this work, we hold the hopping parameter fixed at its critical
value, $\kappa_c=0.156911$,\cite{XLFscaling} leaving the quark mass directly
proportional to $\mu$.  Our pion form factor studies are
carried out with $\mu=0.030$ and $\mu=0.015$.  We also choose the
temporal separation of the source and sink to be $t_f-t_i=15$ time steps.
The additional cases of
$\mu=0.007$, 0.003 and 0.001 are used below as insightful exercises for
GMRES-DR.  Periodic boundary conditions are used in all directions.

Frezzotti and Rossi have shown that, when the hopping parameter is set to its
critical value (so-called ``maximal twist''), masses
and correlation functions with vanishing spatial momenta
are automatically $O(a)$ improved in tmQCD.\cite{FreRos}
A generic matrix element with non-zero spatial
momenta can be improved by averaging over momenta of equal magnitude but
opposite sign as follows,
\begin{equation}
\left<f,\vec{k}|B|i,\vec p\right>+\eta_{i,f,B}\left<f,-\vec{k}|B|i,-\vec p
\right> =
2\xi_B\left<f,\vec{k}|B|i,\vec p\right>|_{{\rm continuum}}+O(a^2)
\label{mom-ave-matrix-element}
\end{equation}
where $\eta_{i,f,B}=\pm 1$ is an overall parity (see Ref.~\cite{FreRos} for
the precise definition of this parity) for the matrix element between
the initial state $|i,\vec p\rangle$,
the final state $|f,\vec{k}\rangle$, and the operator $B$.
The renormalization coeffecient $\xi_B$ relates the continuum and lattice
operators.
Since the energies obtained from a two-point correlator depend only on
$|\vec{k}|^2$ these energies,
like masses, are automatically improved without momentum averaging.

\section{Matrix inversion}\label{sec:gmresdr}

Some of the standard matrix inverters used in lattice QCD research, such as
the stabilized biconjugate gradient, fail
when applied to tmQCD at maximal twist for sufficiently light
quarks.\cite{inverters}
Fortunately there are other inversion algorithms that succeed for
tmQCD inversions, such as conjugate gradient, conjugate gradient squared
and GMRES.\cite{inverters}  The present work made use of the
GMRES-DR algorithm\cite{GMRESDR}, and the remainder of this section contains
some information about our experience with this inverter.

The GMRES-DR inverter is built on the standard GMRES (generalized
minimal residual) matrix inverter, but extends it to incorporate deflation
(D) of the
smallest eigenvalues even after subsequent restarts (R) of the basic
GMRES algorithm.  Since GMRES-DR is a significant improvement over standard
GMRES, and since standard GMRES can successfully invert tmQCD matrices,
it is interesting to explore the application of GMRES-DR to tmQCD.

GMRES uses a Krylov vector space of some dimensionality (let's call it $n$)
chosen by the user and GMRES-DR identifies and retains the $k$-dimensional
subspace spanned by light eigenvectors, where $k$ is also chosen by the user.
For the present work, $n$ and $k$ were chosen to minimize the wall clock time
needed to reach a residual of $|r|<10^{-6}$, where $r\equiv b-Mx$ for Dirac
matrix $M$ and source vector $b$.  This optimization was done at
$\kappa=\kappa_c$ and $\mu=0.030$ and
for our implementation of GMRES-DR($n$,$k$) the result was $(n,k)=(40,10)$.

For our ensemble of 100 configurations, all GMRES-DR(40,10) inversions were
successful at $\mu=0.030$, 0.015 and 0.007.  The pion form factor was not
computed at $\mu=0.007$ due to the onset of finite volume effects, but the
pseudoscalar two-point correlator was computed for $\mu\leq0.007$ as a means
of gaining some experience with GMRES-DR.
At $\mu=0.003$, GMRES-DR(40,10) failed to compute one column out of 1200
but increasing the Krylov subspace to GMRES-DR(60,10) brought success.
At $\mu=0.001$, GMRES-DR(40,10) failed to compute three columns out of 1200
but GMRES-DR(60,10) was again completely successful.
Recall that our choice of $(n,k)=(40,10)$ arose from optimization at
$\mu=0.030$; we did not optimize separately at these very small $\mu$ values.

\begin{figure}[hbt]
\includegraphics[width=12cm]{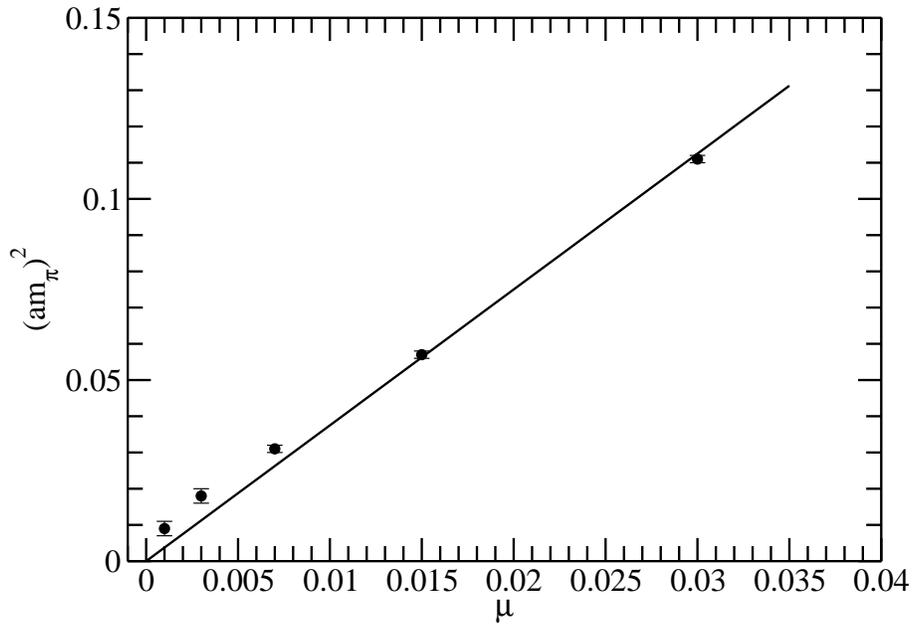}
\caption{Pseudoscalar mass versus the twisted quark mass parameter $\mu$.
         The solid curve is a fit of the data at $\mu=0.030$ and $0.015$ to
         a straight line through the origin.  Data points below $\mu=0.007$
         were computed for the exploration of GMRES-DR, not for
         phenomenological use.}
\label{scaling-of-mpi}
\end{figure}

Table~\ref{tab:GMRESDR} displays the average number of matrix-vector products
that were computed to obtain one column of the inverse to a residual of
$|r|<10^{-6}$ using GMRES-DR(40,10).
Since this number of matrix-vector products depends
on our particular source (i.e. a point source with
specific color index and Dirac index) and also on
our particular initial value for the solution vector,
it is more useful to report the change in $|r|$ relative
to its initial (i.e. before any GMRES-DR iterations) value.
In the present case, the initial residual was
$|r_0|=40.62$ so the data in Table \ref{tab:GMRESDR} represent
the number of matrix-vector products computed to reach
$|r/r_0|<2.5\times10^{-8}$.  This is the quantity that can
be meaningfully compared to studies with other source
vectors.\cite{multiRHS}
Figure~\ref{scaling-of-mpi} shows the pseudoscalar mass squared as a function
of $\mu$ as well as the result of fitting the two largest $\mu$ data points to
a straight line through the origin.  Finite volume effects are apparent for
$\mu\leq0.007$.

\section{Results}
We first analyze the pseudoscalar and vector two-point correlators.
Energies are obtained by fitting the pseudoscalar correlators to
Eq.~(\ref{2pt-decomposition-periodic}) and the vector correlators to the
analogous expression.
Spatial components of nonzero momenta are averaged over all spatial
directions to improve statistics; the three spatial components of the
vector operator are also averaged.

\begin{figure}[htbp]
\includegraphics[width=12cm]{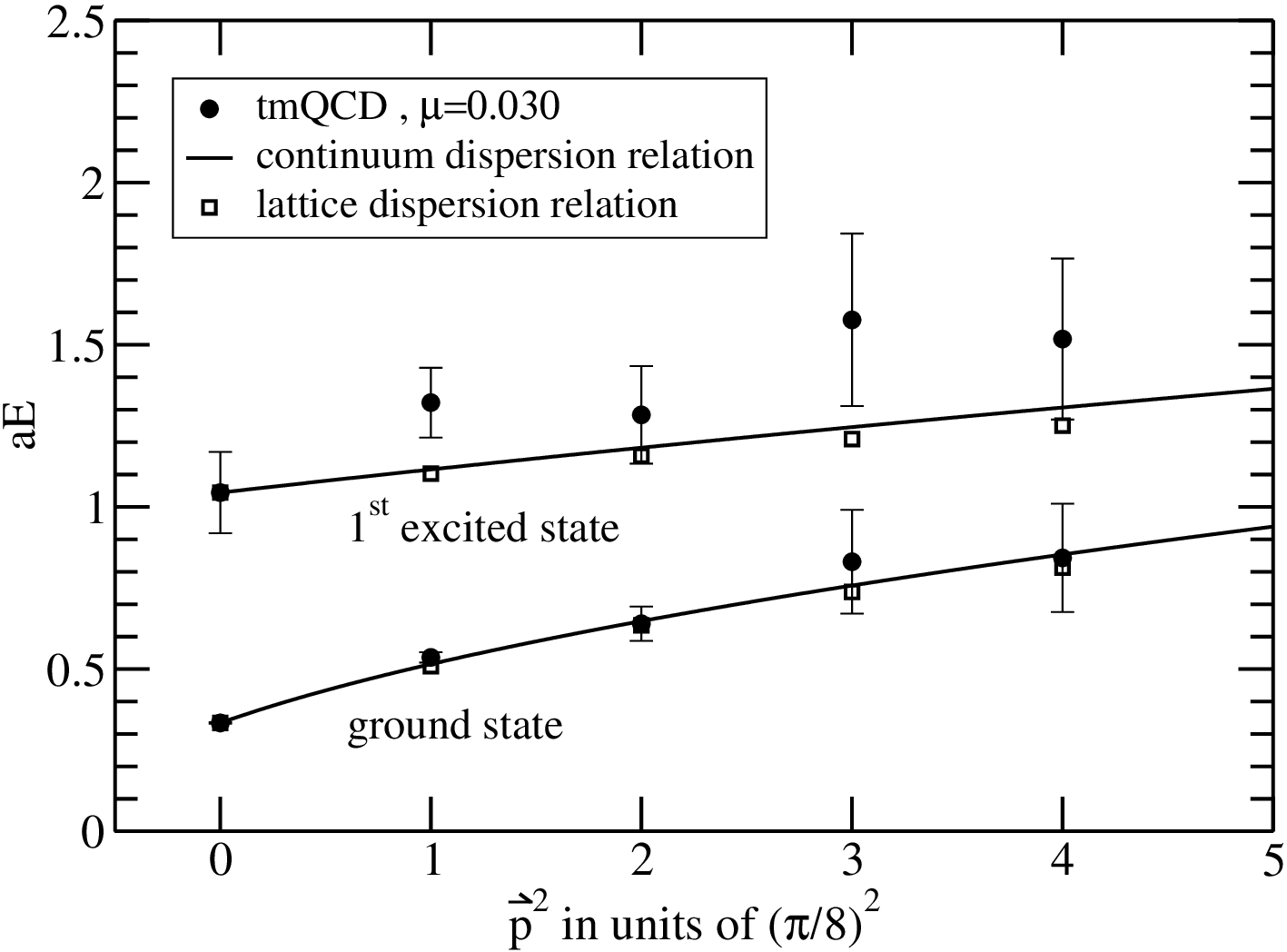}
\caption{Ground state and first excited state pseudoscalar meson energies from
three-state fits to tmQCD at $\mu=0.030$, as compared to the continuum and
lattice dispersion relations.}
\label{pse-mu030-dispersion}
\end{figure}

\begin{figure}[htbp]
\vspace{8mm}
\includegraphics[width=12cm]{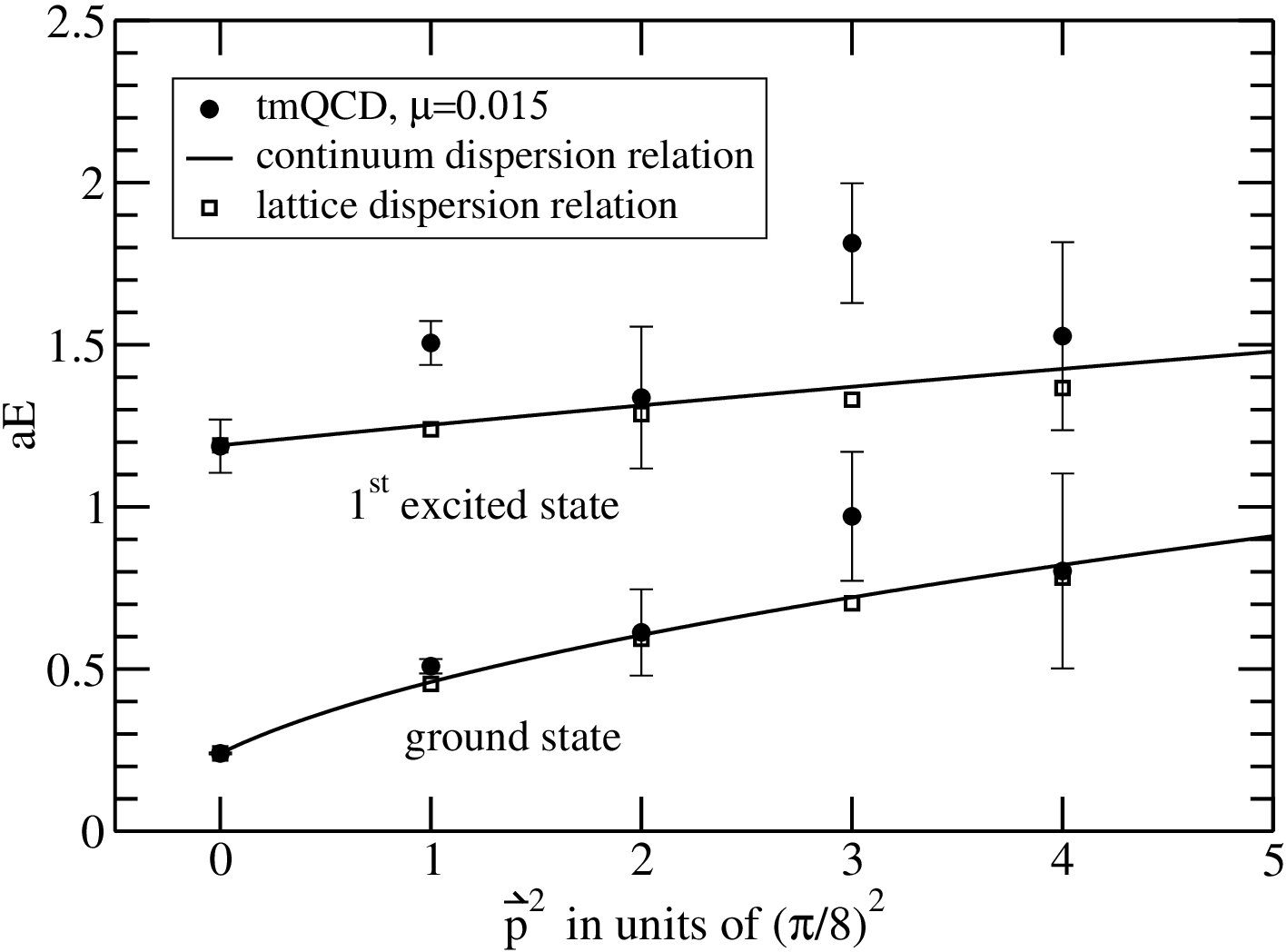}
\caption{Ground state and first excited state pseudoscalar meson energies from
three-state fits to tmQCD at $\mu=0.015$, as compared to the continuum and
lattice dispersion relations.}
\label{pse-mu015-dispersion}
\end{figure}

\begin{figure}[htbp]
\includegraphics[width=12cm]{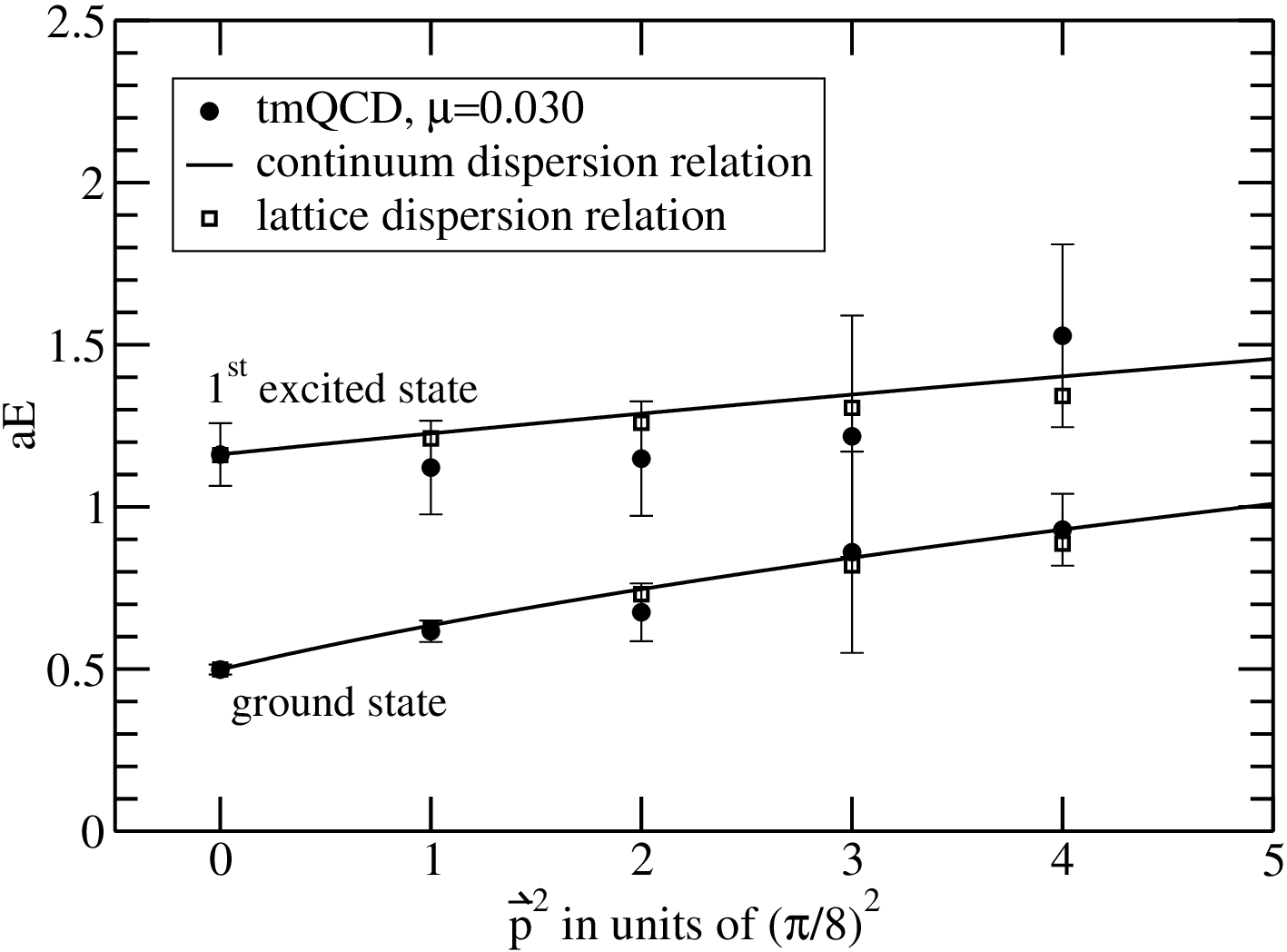}
\caption{Ground state and first excited state vector meson energies from
three-state fits to tmQCD at $\mu=0.030$, as compared to the continuum and
lattice dispersion relations.}
\label{vec-mu030-dispersion}
\end{figure}

\begin{figure}[htbp]
\vspace{8mm}
\includegraphics[width=12cm]{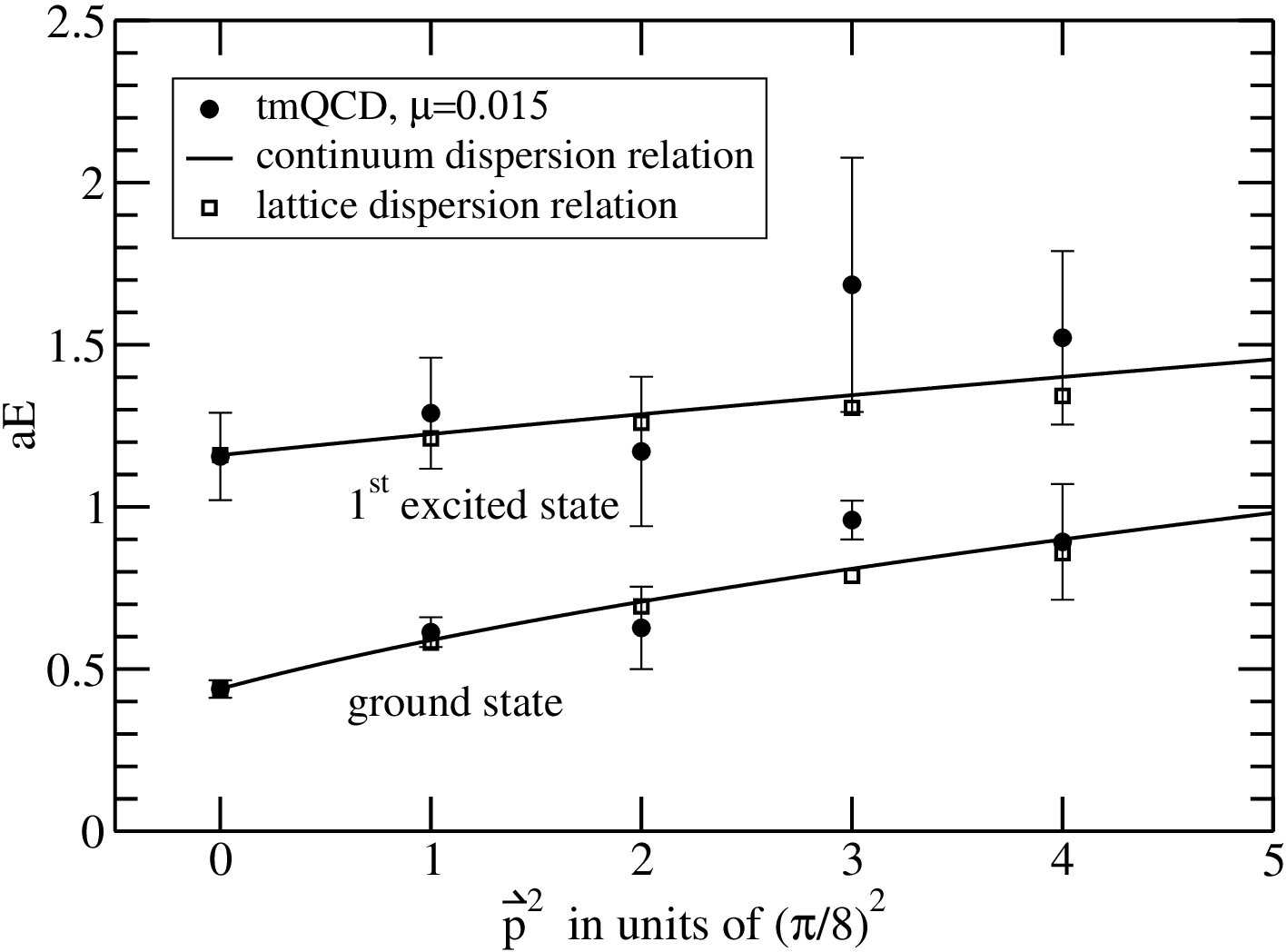}
\caption{Ground state and first excited state vector meson energies from
three-state fits to tmQCD at $\mu=0.015$, as compared to the continuum and
lattice dispersion relations.}
\label{vec-mu015-dispersion}
\end{figure}

Single state fits to the data show a convincing ground state signal for
$|\vec{p}|^2\le4p_{{\rm min}}^2$, where $p_{{\rm min}}=2\pi/L$ and
$L=16$ is the spatial size of our lattice. 
Multi-state fits were also performed, and led to consistent results for the
ground state energies.
These results can be compared to the predictions of the continuum and lattice
dispersion relations given by
\begin{eqnarray}
(aE_{cont.})^2 &=& (aM)^2+|\vec{p}|^2, \label{contdisp} \\
\sinh^2\left(\frac{aE_{latt.}}{2}\right)&=& \sinh^2\left(\frac{aM}{2}\right)+
\sum_{i=1}^3\sin^2\left(\frac{p_i}{2}\right), \label{latdisp}
\end{eqnarray}
respectively.
Figures \ref{pse-mu030-dispersion}, \ref{pse-mu015-dispersion},
\ref{vec-mu030-dispersion} and \ref{vec-mu015-dispersion} show this
comparison where the mass parameters in Eqs.~(\ref{contdisp}) and
(\ref{latdisp}) were chosen to match the lattice data at $\vec p=\vec 0$.
Table \ref{pse-and-vec-masses} contains the numerical values of the pion and
$\rho$ meson masses at those two values of $\mu$ for which the pion form
factor is calculated.

To extract the pion form factor, we performed a simultaneous fit over 
the pseudoscalar two-point correlator with momentum $\vec p_i$,
the pseudoscalar two-point correlator with momentum $\vec p_f$,
and the pseudoscalar($\vec p_i$)-vector-pseudoscalar($\vec p_f$) three-point
correlator.
The fourth component of the conserved vector current was used.
To verify the stability of the ground state, we've performed both a single
state fit over the large time
ranges of the two-point and three-point correlators where the ground state
pion dominates using Eqs.~(\ref{asymptotic-correlators1}) and
(\ref{asymptotic-correlators2}),
and a three state fit over the entire time range (except the source time step)
where the ground state pion as well as first and second excited states are
included.
This latter method involves $3^2=9$ form factors, $F_{\alpha\beta}$, from
Eq.~(\ref{transition}).
For clarity, here are the explicit forms of the correlators used
for the three state fit:
\begin{equation}
G_{\pi\pi}^{initial}(t_i,t,\vec{p}_i) = 
\sum_{n=1}^{3} \frac{Z_n^i{Z_n^i}^*}{E_n(\vec{p}_i)}e^{-\left(\frac{N_t}{2}
\right)E_n(\vec{p}_i)}
\cosh\left[\left(t-t_i-\frac{N_t}{2}\right)E_n(\vec{p}_i)\right],
\label{initial-2pt-decomposition-periodic-3state}
\end{equation}
\begin{equation}
G_{\pi\pi}^{final}(t_i,t,\vec{p}_f) = 
\sum_{n=1}^{3} \frac{Z_n^f{Z_n^f}^*}{E_n(\vec{p}_f)}e^{-\left(\frac{N_t}{2}
\right)E_n(\vec{p}_f)}
\cosh\left[\left(t-t_i-\frac{N_t}{2}\right)E_n(\vec{p}_f)\right],
\label{final-2pt-decomposition-periodic-3state}
\end{equation}
\begin{equation}
\Gamma_{\pi\mu\pi}(t_i,t,t_f,\vec p_i,\vec p_f)=\sum_{n=1}^{3} \sum_{m=1}^{3} Z_m^f
{{e^{-(t_f-t)E_m(\vec{p}_f)}}\over{2E_m(\vec{p}_f)}}
\left<\pi_m(\vec{p}_f)|j_\mu(0)|\pi_n(\vec{p}_i)\right> 
{{e^{-(t-t_i)E_n(\vec{p}_i)}}\over{2E_n(\vec{p}_i)}}{Z_n^i}^*.
\label{3pt-decomposition-3state}
\end{equation}
As will be shown, the ground state is
quite stable regardless of whether the single state fit or three state fit is
used.
The intermediate case of fitting to a ground state plus one
excited state leads to similar results, and will not be discussed further.

For the single state fit, the fitting parameters are the
energies $E$, the prefactors $Z$, and the form factor $F(Q^2)$ from
Eqs.~(\ref{asymptotic-correlators1}) and (\ref{asymptotic-correlators2}).
For the three state fit, the fitting parameters are the
energies $E$ and prefactors $Z$ from
Eqs.~(\ref{initial-2pt-decomposition-periodic-3state}),
(\ref{final-2pt-decomposition-periodic-3state})
and (\ref{3pt-decomposition-3state}) as well as
the form factors $F(Q^2)$ and $F_{\alpha\beta}$ from
Eqs.~(\ref{piff-matrix-element}) and (\ref{transition}).
A standard unconstrained $\chi^2$ minimization fitting procedure was used.
Results from a single state fit to the large (Euclidean) time region were
found to be consistent with results from a three state fit beginning just
one time step from the source.

\begin{figure}[htbp]
\includegraphics[width=10cm]{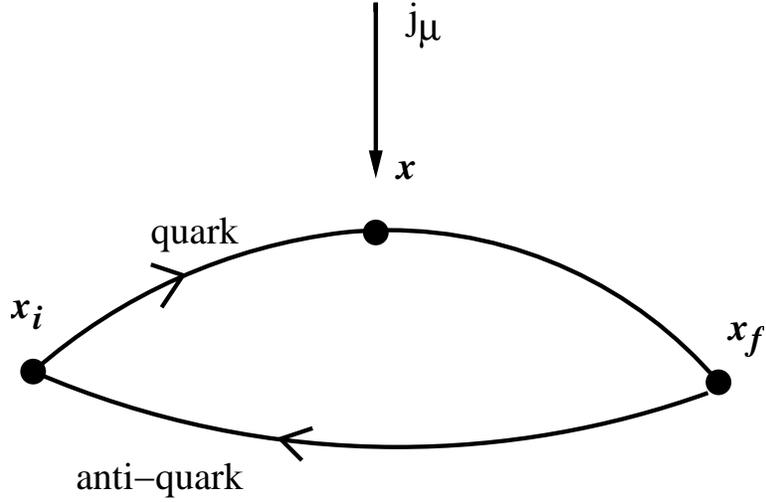}
\caption{Three-point correlator for the pion form factor.}
\label{3point}
\end{figure}

Various choices for $\vec p_i$ and $\vec p_f$ correspond to comparable
momentum transfers, $Q^2$, but the cleanest data come when $|\vec p_i|$ and
$|\vec p_f|$ are minimal.  It is therefore not desirable to restrict
oneself exclusively to $\vec p_f=\vec 0$.  For $\mu=0.015$ we have computed
with $\vec p_f = (0,0,\pm p_{min})$ as well as $\vec p_f=\vec 0$.  For
$\mu=0.030$ we only computed with $\vec p_f = (0,0,\pm p_{min})$.
Each new value of $\vec p_f$ required a new matrix inversion, since the
calculation was performed by combining the propagator from $x_i$ to $x$ with
the sequential propagator from $x_i$ to $x_f$ to $x$.
(See Fig.~\ref{3point}.)
All values of $\vec p_i$ that produce the same $\vec q$ were averaged.
Finally, the momentum averaging procedure of Eq.~(\ref{mom-ave-matrix-element})
was employed to remove $O(a)$ errors.

\begin{figure}[htbp]
\includegraphics[width=12cm]{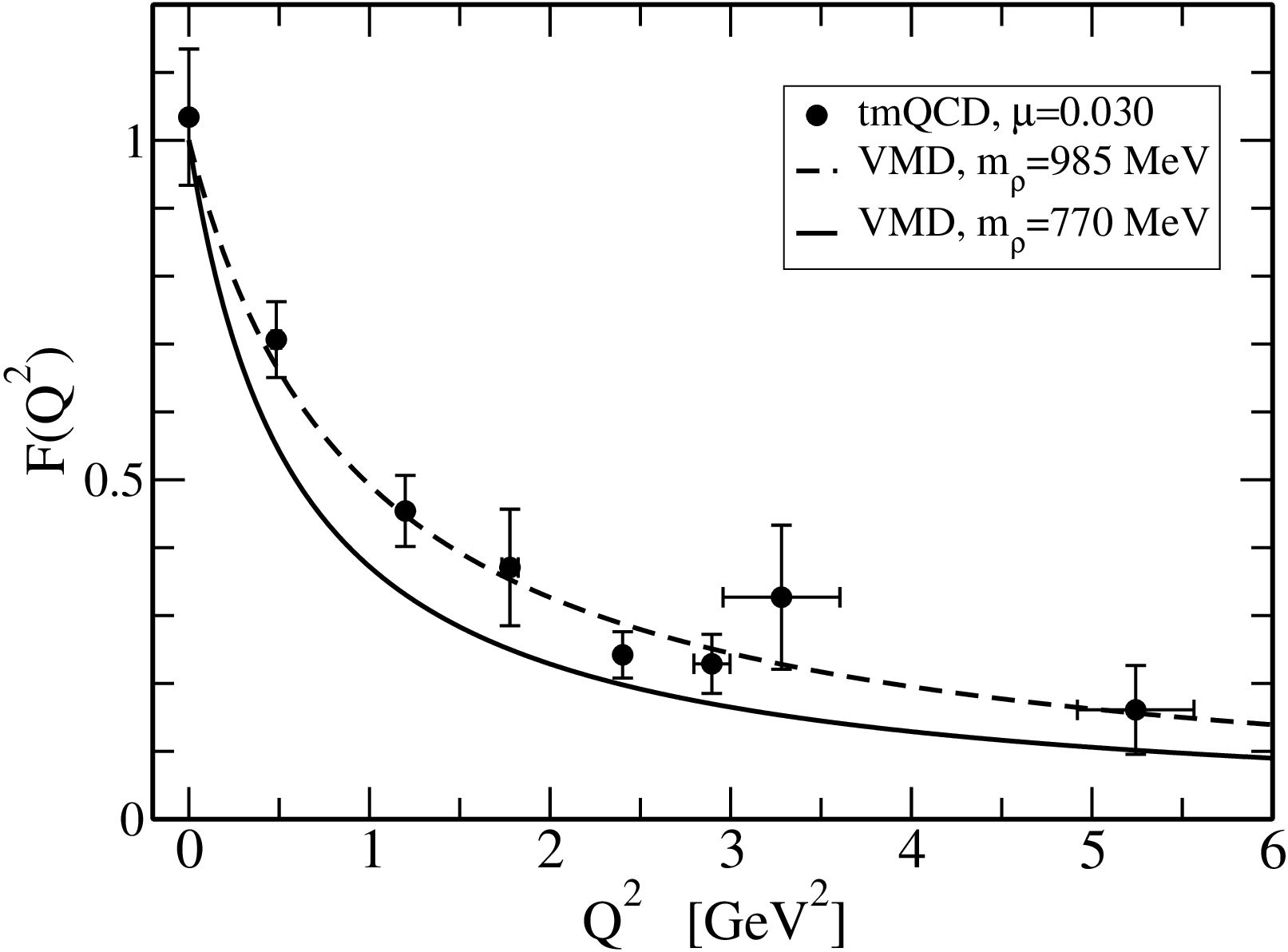}
\caption{The form factor at $\mu=0.030$ obtained from single state fitting as 
compared to vector meson dominance.}
\label{FF-030-1exp}
\end{figure}

\begin{figure}[htbp]
\includegraphics[width=12cm]{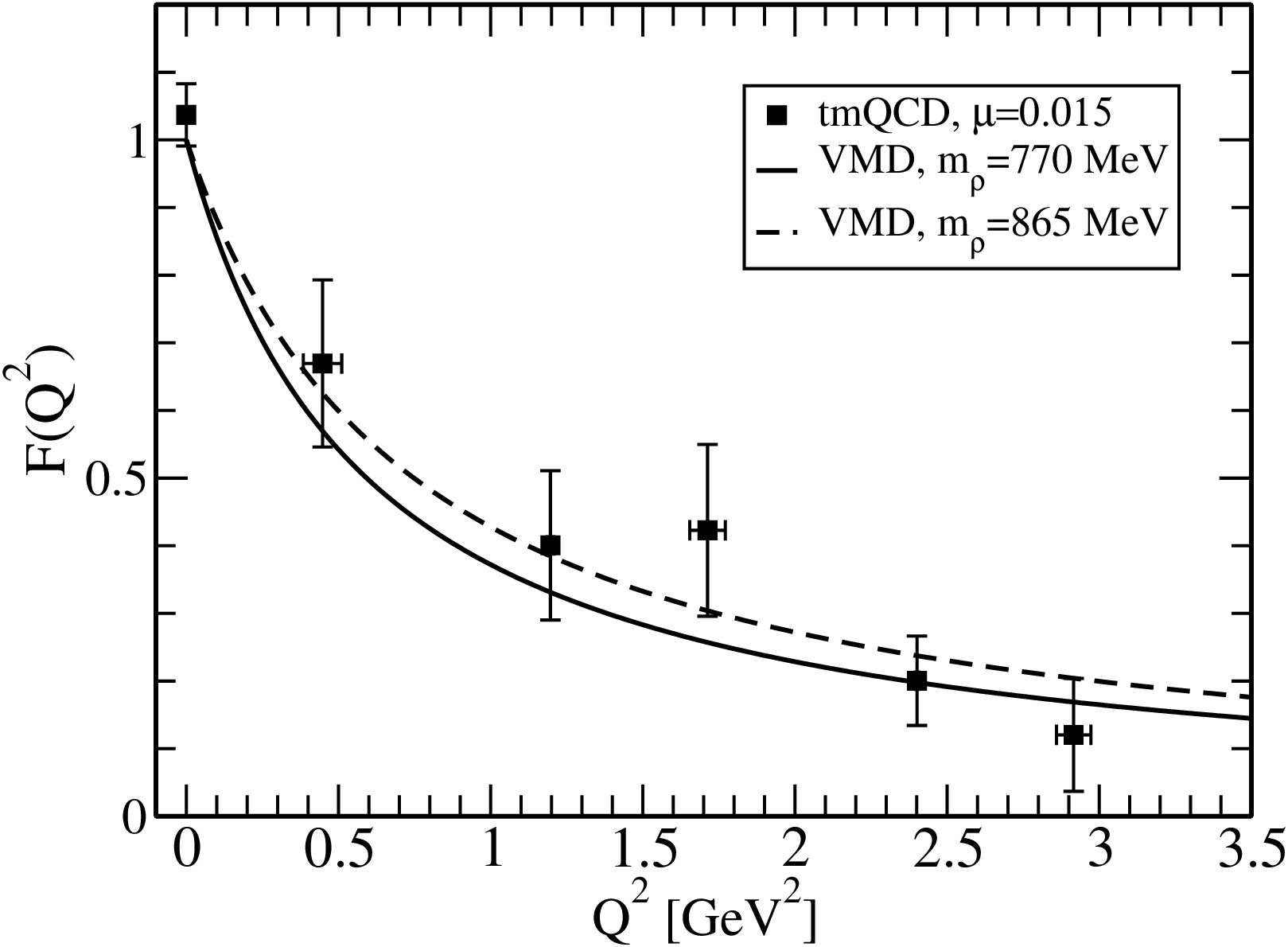}
\caption{The form factor at $\mu=0.015$ obtained from single state fitting as 
compared to vector meson dominance.}
\label{FF-015-1exp}
\end{figure}

\begin{figure}[htbp]
\includegraphics[width=12cm]{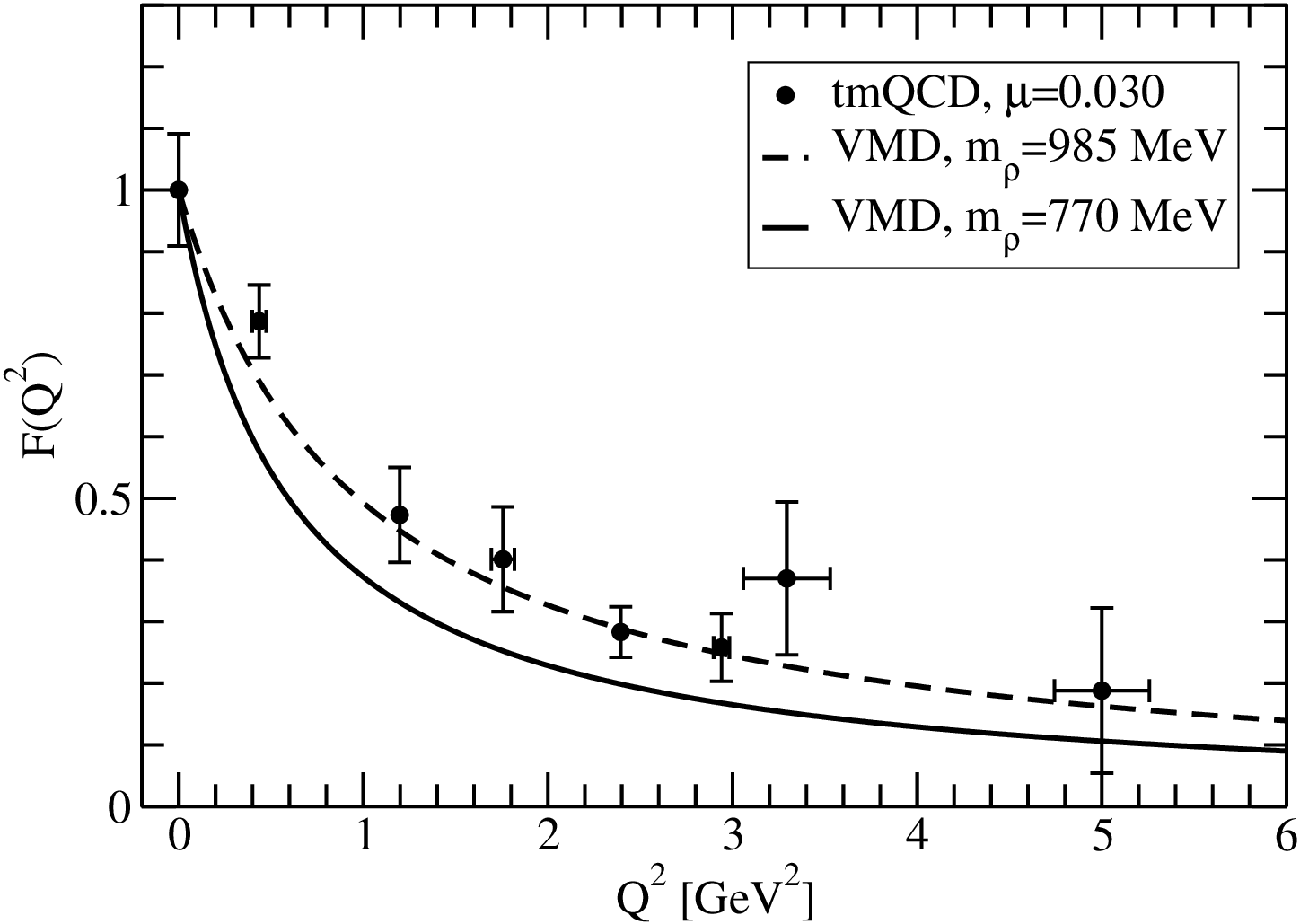}
\caption{The form factor at $\mu=0.030$ obtained from three state fitting as 
compared to vector meson dominance.}
\label{FF-030-3exp}
\end{figure}

\begin{figure}[htbp]
\includegraphics[width=12cm]{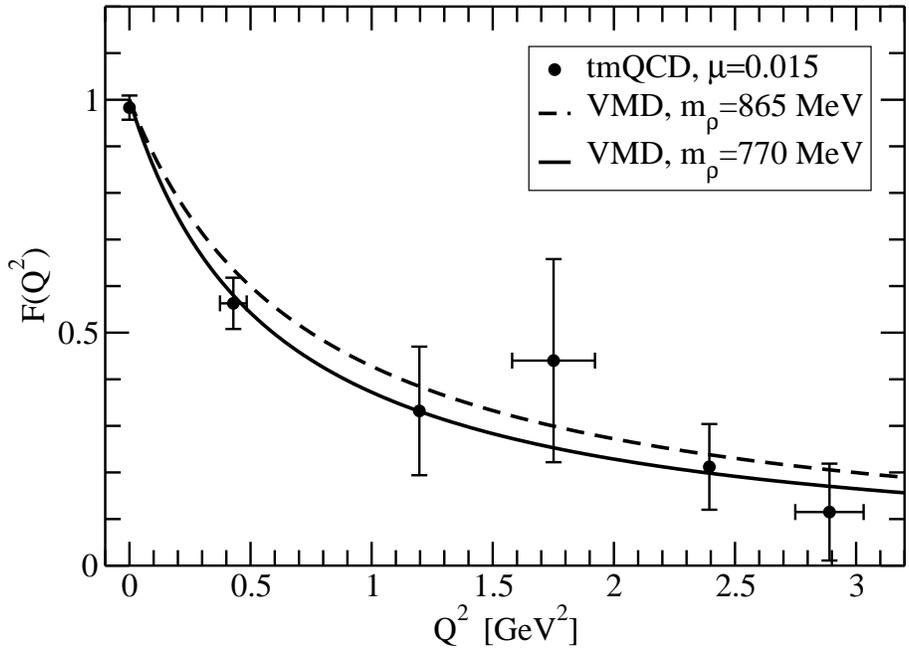}
\caption{The form factor at $\mu=0.015$ obtained from three state fitting as 
compared to vector meson dominance.}
\label{FF-015-3exp}
\end{figure}

In Tables \ref{tab:FF-030-1exp} and \ref{tab:FF-015-1exp}
we show the results for the
pion form factor obtained from one state fits at $\mu=0.030$ and $0.015$
respectively.  These same data are displayed graphically in
Figs.~\ref{FF-030-1exp} and \ref{FF-015-1exp}.
The physical scale was set to $a=0.10$ fm\cite{alat}.
The data show agreement 
with the corresponding vector meson dominance curves.
Results from three state fitting are completely consistent with the one
state fits, as shown in Tables \ref{tab:FF-030-3exp} and \ref{tab:FF-015-3exp}
and Figs.~\ref{FF-030-3exp} and \ref{FF-015-3exp}.
Notice the advantage of using a conserved current: the normalization of the
form factor is $F(0)=1$ without any multiplicative renormalization factor
$Z_V$.

\begin{figure}[hbtp]
\includegraphics[width=12cm]{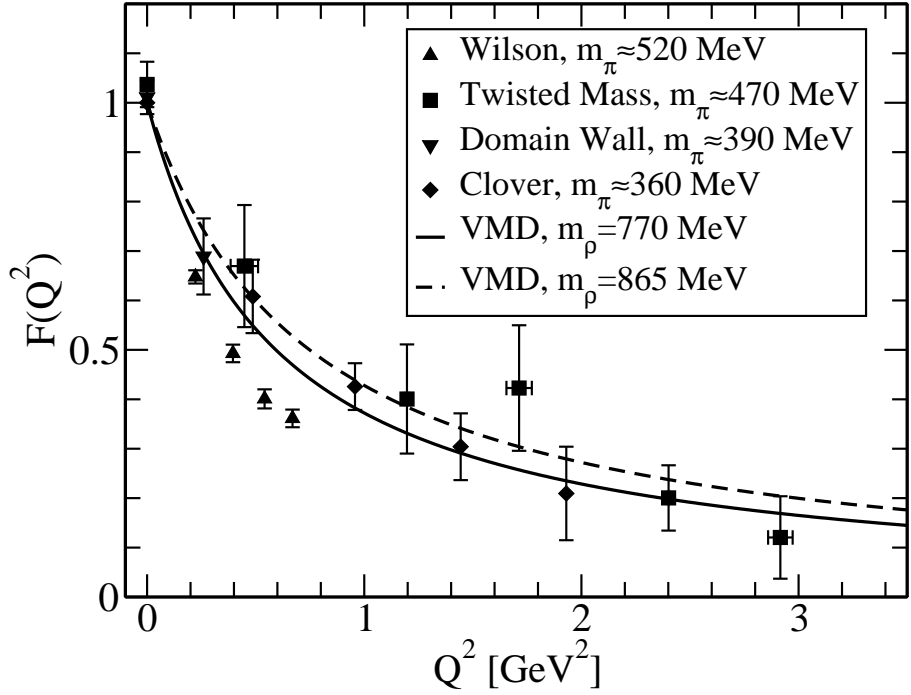}
\caption{Comparing the quenched tmQCD pion form factor with other quenched
         lattice calculations (Wilson from Ref.~\protect\cite{LHPC},
         clover from Ref.~\protect\cite{vanderH},
         domain wall from Ref.~\protect\cite{Nemoto})
         and with vector meson dominance.}
\label{FF-compare}
\end{figure}

In Fig.~\ref{FF-compare}, we compare our results at $\mu=0.015$ to
other quenched calculations of the pion form factor at similar quark masses.
Two vector meson dominance (VMD) curves are also shown --- one with the
physical $\rho$ meson mass and the other with the vector meson mass taken
from our tmQCD simulation at $\mu=0.015$.  The available
experimental data are known to follow VMD with the physical $\rho$ meson mass.
As evident from Fig.~\ref{FF-compare}, the Wilson results have a large
systematic lattice discretization error while the tmQCD
results are consistent with other improved action results and with VMD.
Since the quark mass in the Wilson simulation of Fig.~\ref{FF-compare} is
somewhat larger than the quark masses in the other simulations, one would
expect the Wilson form factor to be slightly {\em larger} than the others.
The plot clearly shows the opposite effect, suggesting that the $O(a)$
contributions are large.  As discussed in Ref.~\cite{LHPC} the apparent
smallness of the Wilson form factor is correlated with the smallness of the
Wilson vector meson mass\cite{EdwardsHK}.  For example,
the Wilson form factor would be consistent with VMD if the physical scale
(i.e. the lattice spacing needed to compute values of $Q^2$ for the
horizontal axis in Fig.~\ref{FF-compare}) were obtained from the
vector meson mass itself.
The tmQCD action does not have large $O(a)$ contributions to the vector
meson mass\cite{XLFscaling} and we find correspondingly small $O(a)$
contributions to the pion form factor.

\begin{figure}[hbt]
\includegraphics[width=12cm]{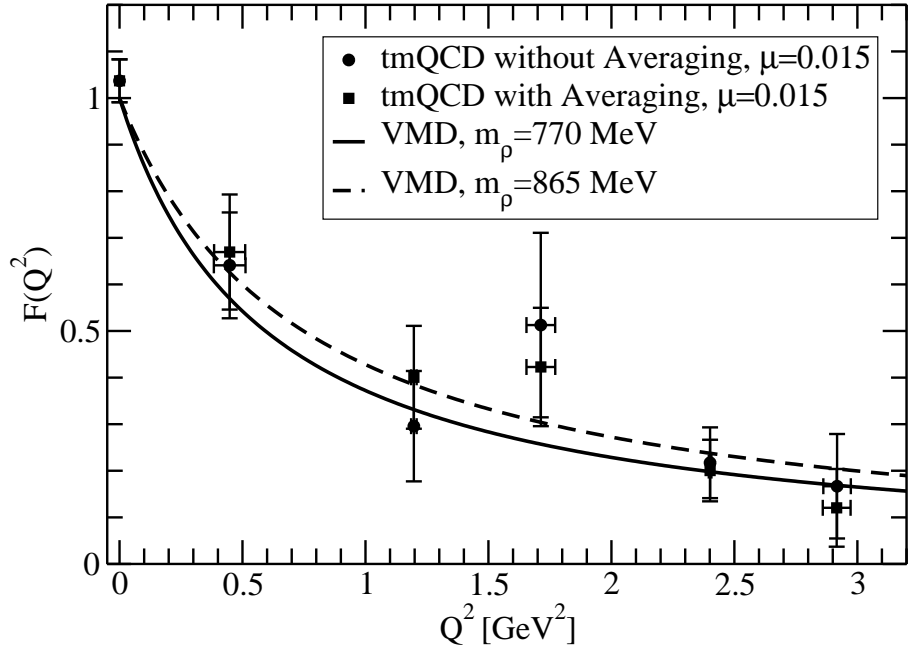}
\caption{The pion form factor from tmQCD at $\mu=0.015$ with and without
         momentum averaging.}
\label{effect-of-averaging}
\end{figure}

As discussed above, the tmQCD results are improved through momentum averaging
at maximal twist ($\kappa=\kappa_c$).
Unimproved tmQCD results would be obtained simply by
omitting the momentum averaging step.  Figure \ref{effect-of-averaging}
shows the tmQCD pion form factor results at $\mu=0.015$ obtained from a
one state fit with and without the averaging procedure.  Interestingly, the
data are quite consistent within the statistical
uncertainties of our simulation.
 
\acknowledgments

The authors thank Walter Wilcox for helpful communications and for
a careful reading of the manuscript.  This work was supported in part by the
Natural Sciences and Engineering Research Council of Canada, the Canada
Foundation for Innovation, the Canada Research Chairs Program and the
Government of Saskatchewan.  Some of the computing was done with WestGrid
facilities.

\begin{table}[hbt]
\caption{The average number of matrix-vector products required to compute one
column of the inverse to a residual of $|r|<(2.5\times10^{-8})|r_0|$ using
GMRES-DR(40,10), where $r_0$ denotes the initial residual.}\label{tab:GMRESDR}
\begin{tabular}{cc}
\hline
$\mu$ & average number of products \\
\hline
0.030 & 816 \\
0.015 & 1175 \\
0.007 & 1384 \\
0.003 & 1450 \\
0.001 & 1489 \\
\hline
\end{tabular}
\end{table}

\begin{table}[tbh]
\caption{Ground state pseudoscalar and vector meson masses.}
\label{pse-and-vec-masses}
\begin{center}
\begin{tabular}{ccc}
\hline
$\mu$& 0.015 & 0.030 \\
\hline
$am_\pi$ & 0.240(3) & 0.334(2) \\
$am_\rho$ & 0.439(27) & 0.499(15) \\
\hline
\end{tabular}
\end{center}
\vspace{5cm}
\end{table}

\begin{table}
\caption{Momentum components in units of $2\pi/L$ used to extract the
pion form factor at $\mu=0.030$ from a single state fit.  $O(a)$ improvement
has been invoked by averaging over data with $n=1$ and $n=-1$.}
\label{tab:FF-030-1exp}
\begin{tabular}{cccc}
\hline
$\vec{p}_f$ [$\frac{2\pi}{L}$] & $\vec{q}$ [$\frac{2\pi}{L}$]
 & $Q^2$ [${\rm GeV^2}$] & $F(Q^2)$ \\
\hline
(0,0,$n$) & (0,0,0) & 0 & $1.034\pm0.100$ \\
(0,0,$n$) & (0,0,$-n$) & $0.484\pm0.025$ & $0.706\pm0.056$ \\
(0,0,$n$) & (0,$\pm1$,$-n$) & $1.198\pm0.007$ & $0.454\pm0.052$ \\
(0,0,$n$) & ($\pm1$,$\pm1$,$-n$) & $1.778\pm0.045$ & $0.370\pm0.086$ \\
(0,0,$n$) & (0,0,$-2n$) & $2.402\pm0$ & $0.242\pm0.034$ \\
(0,0,$n$) & (0,$\pm1$,$-2n$) & $2.896\pm0.1$ & $0.229\pm0.044$ \\
(0,0,$n$) & ($\pm1$,$\pm1$,$-2n$) & $3.281\pm0.323$ & $0.327\pm0.106$ \\
(0,0,$n$) & (0,0,$-3n$) & $5.242\pm0.323$ & $0.161\pm0.065$ \\
\hline
\end{tabular}
\end{table}

\begin{table}
\caption{Momentum components in units of $2\pi/L$ used to extract the
pion form factor at $\mu=0.015$ from a single state fit.  $O(a)$ improvement
has been invoked by averaging over data with $n=1$ and $n=-1$.}
\label{tab:FF-015-1exp}
\begin{tabular}{cccc}
\hline
$\vec{p}_f$ [$\frac{2\pi}{L}$] & $\vec{q}$ [$\frac{2\pi}{L}$]
 & $Q^2$ [${\rm GeV^2}$] & $F(Q^2)$ \\
\hline
(0,0,0) & (0,0,0) & 0 & $1.037\pm0.046$ \\
(0,0,0) & (0,0,$\pm1$) & $0.448\pm0.063$ & $0.669\pm0.124$ \\
(0,0,$n$) & (0,$\pm1$,$-n$) & $1.197\pm0.011$ & $0.401\pm0.110$ \\
(0,0,$n$) & ($\pm1$,$\pm1$,$-n$) & $1.713\pm0.059$ & $0.423\pm0.127$ \\
(0,0,$n$) & (0,0,$-2n$) & $2.401\pm0.005$ & $0.200\pm0.066$ \\
(0,0,$n$) & (0,$\pm1$,$-2n$) & $2.916\pm0.056$ & $0.120\pm0.083$ \\
\hline
\end{tabular}
\end{table}

\begin{table}
\caption{Momentum components in units of $2\pi/L$ used to extract the
pion form factor at $\mu=0.030$ from a three state fit.  $O(a)$ improvement
has been invoked by averaging over data with $n=1$ and $n=-1$.}
\label{tab:FF-030-3exp}
\begin{tabular}{cccc}
\hline
$\vec{p}_f$ [$\frac{2\pi}{L}$] & $\vec{q}$ [$\frac{2\pi}{L}$]
 & $Q^2$ [${\rm GeV^2}$] & $F(Q^2)$ \\
\hline
(0,0,$n$) & (0,0,0) & 0 & $1.0\pm0.091$  \\
(0,0,$n$) & (0,0,$-n$) & $0.436\pm0.038$ & $0.787\pm0.059$ \\
(0,0,$n$) & (0,$\pm1$,$-n$) & $1.197\pm0$ & $0.473\pm0.077$ \\
(0,0,$n$) & ($\pm1$,$\pm1$,$-n$) & $1.756\pm0.063$ & $0.401\pm0.085$ \\
(0,0,$n$) & (0,0,$-2n$) & $2.394\pm0$ & $0.283\pm0.041$ \\
(0,0,$n$) & (0,$\pm1$,$-2n$) & $2.94\pm0.042$ & $0.258\pm0.055$ \\
(0,0,$n$) & ($\pm1$,$\pm1$,$-2n$) & $3.294\pm0.235$ & $0.37\pm0.124$ \\
(0,0,$n$) & (0,0,$-3n$) & $5.0\pm0.257$ & $0.188\pm0.134$ \\
\hline
\end{tabular}
\end{table}

\begin{table}
\caption{Momentum components in units of $2\pi/L$ used to extract the
pion form factor at $\mu=0.015$ from a three state fit.  $O(a)$ improvement
has been invoked by averaging over data with $n=1$ and $n=-1$.}
\label{tab:FF-015-3exp}
\begin{tabular}{cccc}
\hline
$\vec{p}_f$ [$\frac{2\pi}{L}$] & $\vec{q}$ [$\frac{2\pi}{L}$]
 & $Q^2$ [${\rm GeV^2}$] & $F(Q^2)$ \\
\hline
(0,0,0) & (0,0,0) & 0 & $0.983\pm0.026$ \\
(0,0,0) & (0,0,$\pm1$) & $0.429\pm0.055$ & $0.563\pm0.055$ \\
(0,0,$n$) & (0,$\pm1$,$-n$) & $1.197\pm0$ & $0.332\pm0.138$ \\
(0,0,-1) & ($\pm1$,$\pm1$,$-n$) & $1.751\pm0.171$ & $0.44\pm0.218$ \\
(0,0,-1) & (0,0,$-2n$) & $2.394\pm0$ & $0.212\pm0.092$ \\
(0,0,-1) & (0,$\pm1$,$-2n$) & $2.89\pm0.141$ & $0.115\pm0.104$ \\
\hline
\end{tabular}
\end{table}

\end{document}